\documentstyle[12pt]{article}

\newcommand{\ba}{\begin{eqnarray}}
\newcommand{\ea}{\end{eqnarray}}
\begin{document}
\pagestyle{plain}
\title{A relativistic study of the nucleon form factors
\thanks{Partially supported by EC-contract number ERB FMRX-CT96-0008}}
\author{M. De Sanctis $^{a}$, E.~Santopinto$^{b}$, and M.M.~Giannini$^{b}$\\
\and
\begin{tabular}{rl}
$^{a}$&{INFN, Sezione di Roma, P.le A.Moro 2, 00185 Roma},\\
$^{b}$&Dipartimento di Fisica dell'Universit\`a di Genova,\\
      &I.N.F.N., Sezione di Genova\\
      &via Dodecaneso 33, 16164 Genova, Italy\\
      &e-mail:giannini@genova.infn.it\\
\end{tabular}}
\date{}
\maketitle
\noindent
\vspace{4pt}
\begin{abstract}
We perform a calculation of the relativistic corrections to the
electromagnetic elastic form factors of the nucleon obtained with
various Constituent Quark Models. With respect to the non relativistic
calculations a substantial improvement is obtained up to
$Q^2 \simeq 2 (GeV/c)^2$.
\end{abstract}
\begin{center}
PACS numbers: 12.39.Ki, 13.40.Gp, 14.20.Dh
\end{center}

\newpage
\noindent
\section{Introduction}
The non relativistic constituent quark models (CQM) have given good results
in the study of the static properties of the nucleon \cite{is,gia}, like the
baryon spectrum and the magnetic moments, and in a qualitative reproduction
of the photocouplings \cite{cko,ki,aie}. However, the standard CQ-Models are
unable to reproduce the $Q^2$ behaviour of the electromagnetic form factors
even in the low momentum transfer \cite{bil,bff,ck,sig,aie2}.

The use of harmonic oscillator models give rise to form factors which decrease
too fast with respect to the experimental data. Some improvement of this
behaviour, specially in the case of the transition form factors, can be
obtained by using more realistic wave functions \cite{sig,aie}. However, the
problem of a reasonable description of the elastic and transition form factors
of the nucleon in the framework of a Constituent Quark model is still open.

The inclusion of relativistic effects is expected to be important in the
description of the nucleon form factors. The structure of the electromagnetic
current of a relativistic bound system still represents an unsolved problem.
Much attention has been recently devoted to this problem, following
substantially three main lines: the expansion of relativistic current
operators in powers of the inverse quark mass, $\frac{1}{m}$, the evaluation of
the current matrix elements in a light-cone approach and the expansion of the
full relativistic current matrix elements, again in powers of $\frac{1}{m}$.

The first approach takes into account the relativistic effects in the
electromagnetic operators \cite{cl,cap}, the baryon states being the standard
CQM ones. This type of relativistic correction includes the q\={q}-pair
contribution to two-body currents coming from the one-gluon exchange
\cite{bu,per}. The numerical results show that these effects are significant
but not sufficient to explain the data.

There are many interesting results obtained in a light-cone approach
\cite{ck,sal,cot}, such as the fact that the relativistic corrections to the
transition form factors are important  at low $Q^2$ \cite{ck} and that the
root mean
square radius of the proton is increased \cite{ck,cot}. This method is very
useful since it
allows to perform calculations starting from  non relativistic wave functions.
However there are still some problems in extracting form factors from the
evaluated current matrix elements.

In this work we follow the third method which consists of expanding the
current matrix elements in powers of $\frac{1}{m}$ \cite{pfe,dsp} and we
propose a simplified approach useful for a preliminary calculation of the
relativistic corrections to the elastic electromagnetic form factors of the
nucleon obtained starting from different Constituent Quark Models. The use of
Lorentz boosts for the quark spinors ensures that the relation between the
dynamic variables of the
initial and final states is relativistically correct.
On the other hand, we assume
that the quark internal motion is well described
by the standard
non relativistic wave function.
The current matrix elements are constructed with a quark current operator
containing only one-body terms and no quark form factors are introduced.
We point out that
the non relativistic expansion of the matrix elements of the present work,
up to order
$m^{-2}$, is coincident with that given by standard procedures
\cite{friar, kf,dsp} introduced for the few-nucleon systems
and no approximation is done with
respect to the momentum transfer $Q^2$ dependence.

In Sec. 2, we describe the evaluation of the current matrix elements
arriving at simple analytical expressions for the form factors. In Sec. 3,
we discuss the results obtained with various 3q-wave functions and make a
comparison with the experimental data. A brief conclusion is given in Sec.
4.

\section{The current matrix elements}
For the study of the transition process  between the
initial (I) and final (F) states, we have to calculate the current matrix
element

\begin{equation}
J^{\mu}_{FI} = \langle{\bar{\Psi}}_{F} ~|~\sum_{i=1}^{3}j^{\mu}(i)
~|~\Psi_{I}~\rangle ~,
\end{equation}

\noindent where $j^{\mu}(i)$ is the e.m. current of the i-th quark. We choose
the Breit frame and so the total initial and final tetramomenta $P_I~=~(E_I,
\vec{P}_I)$, $P_F~=~(E_F,\vec{P}_F)$ are related by

\begin{equation}
\vec{P}_{I}~=~-~\vec{P}_{F}~=~ -~\frac{\vec{q}}{2}~,~~~~~~
E_{I}=E_{F}=~\sqrt{M^{2}+\frac{{\vec{q}}^2}{4}}~\equiv~E~
\end{equation}
where $~\vec{q}~$ is the virtual photon momentum, $Q^2~=~\vec{q}^2~~$ and $M$
is the nucleon mass.
We denote with $p_{i}^{*}~(i=1,2,3)$ the quark tetramomenta in the nucleon
rest frame and we introduce the relative three-momenta
\begin{equation}
\vec{p}_{\rho}~=~\frac{1}{\sqrt{2}}(\vec{p_{1}}^{*} - \vec{p_{2}}^{*})~,
~~~~~~
\vec{p}_{\lambda}~=~\frac{1}{\sqrt{6}}(\vec{p_{1}}^{*} + \vec{p_{2}}^{*}-2
\vec{p_{3}}^{*})
\end{equation}
which are conjugated to the standard Jacobi coordinates $\vec{\rho}$ and
$\vec{\lambda}$.
The 3-quark state is assumed to be

\begin{equation}
\Psi_{I}~=~\prod_{i=1}^{3}~B_{i}~u_{i}(p_{i}^{*})~\phi(\vec{p_{\rho}}
,\vec{p_{\lambda}})~,
\end{equation}
where the $B_{i}~~(i=1,2,3)$ are the usual Dirac boost operators that transform
the quark spinors $u_{i}(p_{i}^*)$ from the nucleon rest frame to the Breit
one.
The boosted spinors, $\psi_{i}~=~B_{i}u_{i}(p_{i}^{*})$,
have the covariant normalization
\begin{equation}
\bar{{\psi}}_{i}\psi_{i}~=1~.
\end{equation}
In Eq. (4) $\phi(\vec{p_{\rho}},\vec{p_{\lambda}})$  is the standard
non relativistic 3q-wave function, where for
simplicity we have omitted the spin and isospin variables.
The final state is written in a similar way.
The current operator of the i-th quark, $j_{\mu}(i)$, has the form
\begin{equation}
j_{\mu}(i)~=~\sqrt{\frac{m}{\epsilon~'_{i}}}~\gamma_{i}^{\mu}~
\sqrt{\frac{m}{\epsilon_{i}}},
\end{equation}
where $m$ is the quark mass and $\epsilon~_{i}$ ($\epsilon~'_{i}$) is the
initial (final) quark energy in the Breit frame.
The normalization factors $\sqrt{\frac{m}{\epsilon~'_{i}}}$,
$\sqrt{\frac{m}{\epsilon~_{i}}}$, have been introduced in order to obtain for
the
current matrix elements the correct expantion in powers of $\frac{1}{m}$ (
i.e. coincident with what is usually quoted in the literature) as shown in
ref. \cite{friar}.
The quark energies $\epsilon~_{i}$, $\epsilon~'_{i}$ are then
expressed in terms of the corresponding quantities in the nucleon rest
frame by means of
standard Lorentz transformations.

Finally, we add a factor $2E$ to the matrix element of Eq. (1)
in order to take into account the normalization of the total matrix
element.

Because of the
antisymmetry of the
3q-states, we can substitute $\sum_{i=1}^{3}~j_{\mu}(i)$ with $3~j_{\mu}(3)$.
The interacting quark absorbs the photon threemomentum in the Breit frame and
therefore taking into account the Lorentz boost on the 3-quark, we can
write the momentum conservation as follows:

\begin{equation}
{\vec{p}~'}_{\lambda}  = ~\vec{p}_\lambda - \sqrt{{2\over 3}} {M\over E}
\vec{q}~~,~~~~~~~~{\vec{p}~'}_\rho~ = ~\vec{p}_{\rho}
\end{equation}
where the apices refer to the final momenta.  The resulting expression for
the current matrix element is complicated because of the presence of
non local terms coming from the momentum dependence and the
calculation can be performed numerically.  However, in order to arrive
at a preliminary calculation of the relativistic corrections to the e.m.
current, we introduce some simplified assumptions.

First, consistently
with the use of a non relativistic model for the internal nucleon dynamics
we approximate the quark energies in the nucleon rest frame as $\epsilon_{i}^*
~\simeq m$.
Furthermore, we perform an expansion keeping contributions up to the
first order
in the relative quark momenta, but we treat exactly the
dependence on the momentum transfer $\vec{q}$.  To this end, we introduce
in the current matrix element of Eq. (1)
the variable $\vec{\pi}_{\lambda}$ that is related to $\vec{p}_{\lambda}$ and
$\vec{p}_{\lambda}~'$ in the following way

\begin{equation}
\vec{p}~'_{\lambda}  = \vec{\pi}_{\lambda} - \frac{1}{2}\sqrt{{2\over 3}}
{M\over E} \vec{q}
\end{equation}
\begin{equation}
\vec{p}_\lambda  = \vec{\pi}_\lambda + \frac{1}{2}\sqrt{{2\over 3}}
{M\over E} \vec{q}~.
\end{equation}
From the previous equations one also has
\begin{equation}
\vec{\pi}_\lambda = \frac{1}{2}(\vec{p}~'_{\lambda}+\vec{p}_{\lambda})~.
\end{equation}
We expand the current matrix element of equation (1) by keeping up to the
linear terms in $\vec{\pi_{\lambda}}$ and $\vec{p_\rho}$.  In correspondence,
 the electric and magnetic form
factors have zero- and first-order contributions.

The results  for the zero-th order charge and magnetic form factors can be
given in a simple analytical form
\begin{equation}
G_{E}^{(0)}(Q^2) = {E\over M} (t_S)^2 t_I G_{E}^{nr} (Q^2 {M^2\over E^2})
\end{equation}
\begin{equation}
G_{M}^{(0)}(Q^2) = {E\over M} (t_S)^2 t_I {g_{\sigma}\over 2m} G_{M}^{nr} (
Q^2{M^2\over E^2})~,
\end{equation}
where $G_{E}^{nr}$, and $G_{M}^{nr}$ are the electric and magnetic form
factors as given by the non relativistic quark model.  The quantities $t_S$,
$t_I$  and $g_{\sigma}$
\begin{equation}
t_S = {1\over Mm} \bigl[E\eta_S - {M\over E} {Q^2\over 12}\bigr]~,
\end{equation}
\begin{equation}
t_I = {Mm\over E\eta_I + {M~Q^2\over 6E}}
\end{equation}
\begin{equation}
g_{\sigma} = {2\over 3} + {\eta_I\over M}
\end{equation}
with
\begin{equation}
\eta_S = \bigl[m^2 + {M^2\over 36E^2} Q^2\bigr]^{1/2}~,
\end{equation}
\begin{equation}
\eta_I = \bigl[m^2 + {M^2\over 9E^2} Q^2\bigr]^{1/2}~,
\end{equation}
as multiplicative factors.

The first order contribution to the charge density matrix element is
essentially of spin-orbit nature and a non relativistic
expansion up to order $m^{-2}$ gives the sum of the standard and
the anomalous
spin-orbit terms.  The first order term will be omitted in our
calculations, since
it gives a numerically negligible contribution for nucleon states which,
according to the models we use, are mainly in S-wave.
The first order corrections to the magnetic form factors are of two types,
 spin-orbit like and convective.  The first one can be disregarded for the
same reason quoted above for the charge form factor, while the
convective part
gives in any case a small contribution.

Therefore, within these
approximations, the relativistic corrections introduce  two kinds of
modifications with respect to the non relativistic treatment: a
multiplicative factor coming from the expansion of the quark spinors and
the argument of the non relativistic form factors, i.e. the momentum
transfer squared $Q^{2}$, being replaced by $Q^{2}\frac{M^{2}}{E^{2}}$.

The current
matrix elements must satisfy the current conservation equation
\begin{equation}
q_\mu J^\mu_{FI} =0~,
\end{equation}
and it is satisfied in our case since the Constituent Quark Models we
have used are based
on local interactions.

\section{Results and comparison with experimental data}
The form factors of Eqs. (11) and (12) can be
calculated using as input
the nucleon form factors obtained in a non relativistic quark model. We
present the results for different choices of the quark interaction, namely
the h.o. \cite{is} (Fig. 1), the three-body force hypercentral
potential \cite{pl} (Fig. 2) and an
exactly solvable potential based on a hypercoulomb interaction \cite{sig} (
Fig. 2). All these
models have been used for the description of the spectrum \cite{is,pl,sig} and
of the photocouplings \cite{cko,ki,aie,sig}. The three-body force approach
has allowed also a
systematic analysis of the transition form factors for the excitation of
the baryon resonances \cite{aie2}. All of them contain also a spin dependent
(hyperfine) interaction, which is essential for the description of the
$N-\Delta$ splitting and for the excitation of quite a few resonances. For the
elastic form factors, the configuration mixing coming from the hyperfine
interaction does not produce strong effects, apart from the neutron charge
form factor, and we shall omit it here.

In the h.o. case, the choice of the h.o. parameter $\alpha$
is crucial. There are many different values of $\alpha$ quoted in the
literature, according to the quantities to be fitted \cite{dd}. We
report in Fig. 1 the results obtained with 1) $\alpha~=~0.229~GeV$,
which gives the correct r.m.s. radius of the proton \cite{is,iks} without
the relativistic corrections, and
2) $\alpha~=~0.410~GeV$, which is necessary in order to reproduce the
photoexcitation of the $D_{13}$ and $F_{15}$ resonances \cite{cko,ki} and
corresponds to a confinement radius of the order of $0.5~fm$. We note
that the relativistic corrections increase the r.m.s. with respect to the
non relativistic calculation \cite{mds}. In fact from Eq. (11) one gets, for
the h.o. proton charge form factor,
\begin{equation}
<r^2>~=~\frac{1}{\alpha^2}~+~\frac{6}{M^2}~,
\end{equation}
where $M$ is the nucleon mass and $m~=~\frac{M}{3}$. In order to get the
correct radius one
should use $\alpha~=~0.285~GeV$, which however is not too different from
the choice 1).
The results of Fig. 1 show that the relativistic corrections improve the h.o.
form factors, but the $Q^2$ behaviour is still different from the
experimental data.

In Fig. 2 we give the form factors obtained starting from the non relativistic
calculation performed with the three-body force potential of ref. \cite{pl}.
This
potential has the form $V(x)~=~-\frac{\tau}{x}~+~b_{conf}~x$, where $x$ is the
hyperradius $x~=~\sqrt{{\rho}^2~+~{\lambda}^2}$ and the values of the
parameters are
$\tau ~=~4.59$ and $b_{conf}~=~1.61~fm^{-2}$. It should be noted that with
this choice of the parameters and the inclusion of the standard hyperfine
interaction, the three-body force allows
to describe consistently the non-strange baryon spectrum \cite{pl},
the photocouplings \cite{aie} and the electromagnetic transition form factors
\cite{aie2}.

In Fig. 2 we give also the results for the solvable model of ref. \cite{sig}.
It is based on the hypercoulomb potential $V_{hyc}(x)~=~-\frac{\tau}{x}$ ,
with $\tau~=~6.39$, to which a small confinement term is added. The
advantage of this potential is that the results can be given in analytical
form. For instance, the proton charge form factor is given by
\begin{equation}
G_{E}^{nr}(Q^2)~=~\frac{1}{[1~+~\frac{25}{24}~\frac{Q^2}{{\tau}^2~m^2}]^
{\frac{7}{2}}}.
\end{equation}
which, at variance with the h.o. case, for large $Q^2$ has a power-law
behaviour.

From the analysis of the results of Figs. 1 and 2, one sees that in
general the inclusion of relativistic corrections improves significantly the
 non relativistic calculations.
The improvement at low $Q^2$ is related to the correct
non relativistic limit of the current matrix elements. The improvement at
higher $Q^2$ depends
on the
relation between the initial and final state variables and allows to keep
exactly the $Q^2$ dependence of the form
factors.

It should be noted that the simultaneous reproduction of
the spectrum and the photocouplings requires a confinement radius of the
order of $0.5~fm$ \cite{cko,ki,pl,aie}
and the relativistic increase is quite beneficial, but it is still not
sufficient to get nearer to the data. In particular there still remain
problems in
the  $Q^2$ behaviour in the low and medium range. Similar
problems are encountered also in the transition form factors,
both in the relativistic \cite{ck} and in the non relativistic
\cite{aie2, sig} calculations, and so one
can think
that  not only the relativistic corrections are responsible for the
discrepancies between the CQM calculations and the experimental data. As
already noted elsewhere \cite{aie2,sig,bil}, some fundamental dynamical
mechanism (effective at large distance, which means at low $Q^2$)is still
lacking, such as the explicit inclusion of quark-antiquark pairs
both in the baryon states and in the electromagnetic transition operator.

\section{Conclusions}
We have calculated the relativistic corrections to the elastic
nucleon form factors in a simplified and preliminary approach
which leads to simple analytical expressions.
We have used as input different Constituent Quark
Models, namely the harmonic oscillator \cite{is}, the hypercentral model of ref.
\cite{pl} and the analytical model of ref. \cite{sig}  showing that in all
the three models
the relativistic corrections are important, since they
 bring the
non relativistic calculations nearer to the data, but still they are not
sufficient. The persisting discrepancy may be an indication that
further degrees of freedom (q\={q}-pairs, gluons) should be
included in the CQM in a more explicit way.

\noindent
{\bf Acknowledgment}

We gratefully thank  Prof. D.Prosperi for many useful discussions
and suggestions during the development of the present work.

\begin{figure}
\vspace{18cm}
\caption{
The charge (a) and magnetic (b) form factor of the proton and (c) the
magnetic form factor of the neutron. The curves are
the h.o. calculations using $\alpha~=~0.229~GeV$ (dotted and dot-dashed) or
$\alpha~=~0.410~GeV$ (dashed and full). The dashed and dotted curves are
the non relativistic calculations, the full and dot-dashed are the
corresponding relativistic ones, obtained from Eqs. (11) and (12). The
experimental data are taken from the compilation of ref. [15].}
\end{figure}

\begin{figure}
\vspace{18cm}
\caption{
The charge (a) and magnetic (b) form factor of the proton and (c) the
magnetic form factor of the neutron. The curves obtained using the
model of ref. [21] are the non relativistic (dashed) and relativistic
calculations (full).
The curves obtained using the
model of ref. [9] are the non relativistic (dotted) and relativistic
calculations (dot-dashed). The data are the same as in Fig. 1.}
\end{figure}


\begin{thebibliography}{99}


\bibitem{is}
N. Isgur and G. Karl,
Phys. Rev. {\bf D18}, 4187 (1978); {\bf D19}, 2653 (1979).

\bibitem{gia}
M.M. Giannini, Rep. Prog. Phys. {\bf 54}, 453 (1991) and references quoted
therein.

\bibitem{cko}
L. A. Copley, G. Karl and E. Obryk, Phys. Lett. {\bf 29}, 117 (1969).

\bibitem{ki}
R. Koniuk and N. Isgur, Phys. Rev. {\bf D21}, 1868 (1980).

\bibitem{aie}
M. Aiello, M. Ferraris, M.M. Giannini, M. Pizzo and E. Santopinto,
Phys. Lett. {\bf B387}, 215 (1996).

\bibitem{bil}
R.~Bijker, F.~Iachello and A.~Leviatan,
Ann. Phys. (N.Y.) {\bf 236}, 69 (1994).

\bibitem{bff}
R.~Bijker, F.~Iachello and A.~Leviatan,
Phys. Rev. {\bf C54}, 1935 (1996)

\bibitem{ck}
S. Capstick and B.D. Keister, Phys. Rev. {\bf D51}, 3598 (1995)

\bibitem{sig}
E. Santopinto, F. Iachello and M.M. Giannini, Proceedings of the DA$\Phi$CE
Workshop, Frascati, November 11-14, 1996 (Nucl. Phys. A, 1997) and to be
published.

\bibitem{aie2}
M. Aiello, M.M. Giannini and E. Santopinto,
The electromagnetic excitation of negative parity nucleon resonances,
to be published

\bibitem{cl}
F. E. Close and Z. Li, Phys. Rev. {\bf D42}, 2194 (1990).

\bibitem{cap}
S. Capstick, Phys. Rev. {\bf D46}, 2864 (1992).

\bibitem{bu}
A. Buchmann, E. Hernandez and K. Yazaki,
Phys. Lett. {\bf B 269}, 35 (1991); Nucl. Phys. {\bf A 569}, 661 (1994).

\bibitem{per}
E. Perazzi, M. Radici and S. Boffi,
Nucl. Phys. {\bf A 614}, 346 (1997).

\bibitem{sal}
F. Cardarelli, E. Pace, G. Salm\`e and S. Simula,
Phys. Lett. {\bf B 357}, 267 (1995).

\bibitem{cot}
A. Szczepaniak, Ji Chueng-Ryong and S.R.Cotanch,
Phys. Rev. {\bf C 52}, 2738 (1995).

\bibitem{pfe}
M. Warns, H. Schr\"{0}der, W. Pfeil and H. Rollnik,
Z. Phys. {\bf C 45}, 613 (1990).

\bibitem{dsp}
M. De Sanctis and D. Prosperi,
Il Nuovo Cimento {\bf A 98}, 621 (1987) and references quoted therein.

\bibitem{kf}
R.A. Krajcik and L. L. Foldy,
Phys. Rev. {\bf D 12}, 1700 (1975)

\bibitem{friar}
J.L. Friar,
Ann. Phys. (N.Y.){\bf 81}, 332 (1973); Nuc. Phys. {\bf A 264}, 455 (1976).

\bibitem{pl}
M. Ferraris, M.M. Giannini, M. Pizzo, E. Santopinto and L. Tiator, Phys. Lett.
{\bf B364}, 231 (1995).


\bibitem{dd}
D. Drechsel,
Il Nuovo Cimento {\bf 76 A}, 388 (1983).

\bibitem{iks}
N. Isgur, G. Karl and D. W. L. Sprung, Phys. Rev. {\bf D23}, 163 (1981).

\bibitem{mds}
M. De Sanctis,
Il Nuovo Cimento {\bf A 109}, 425 (1996)


\end{thebibliography}
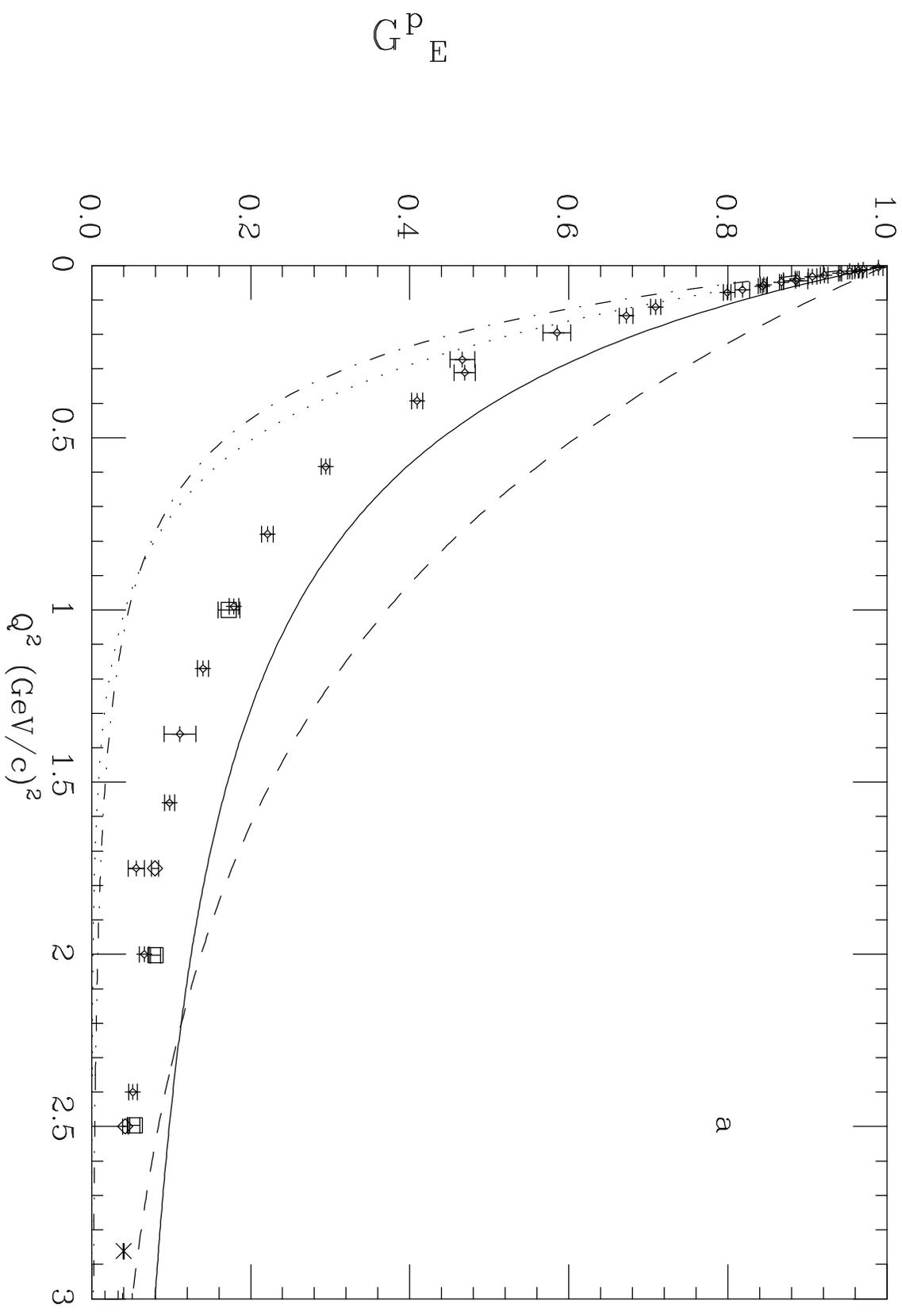
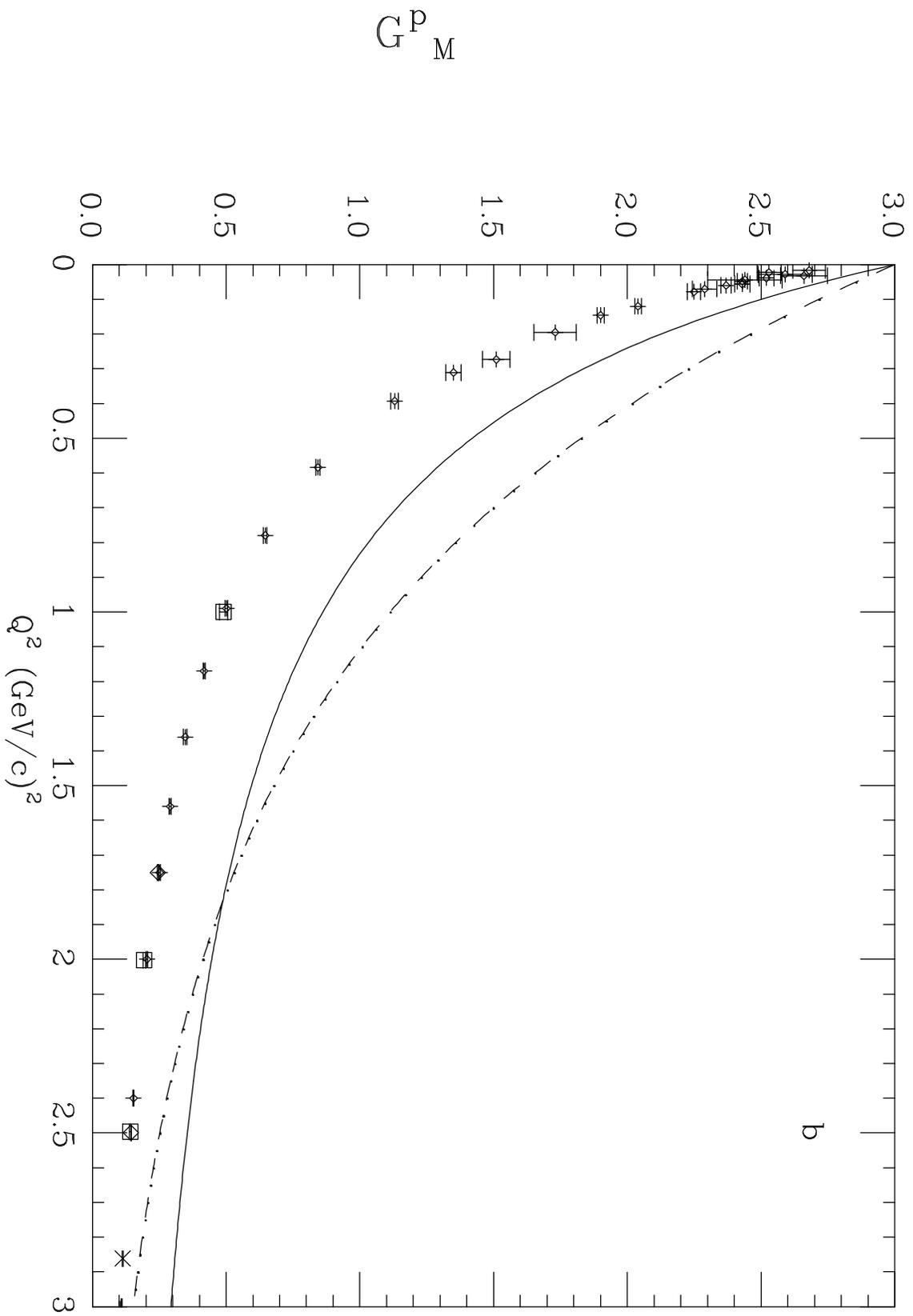
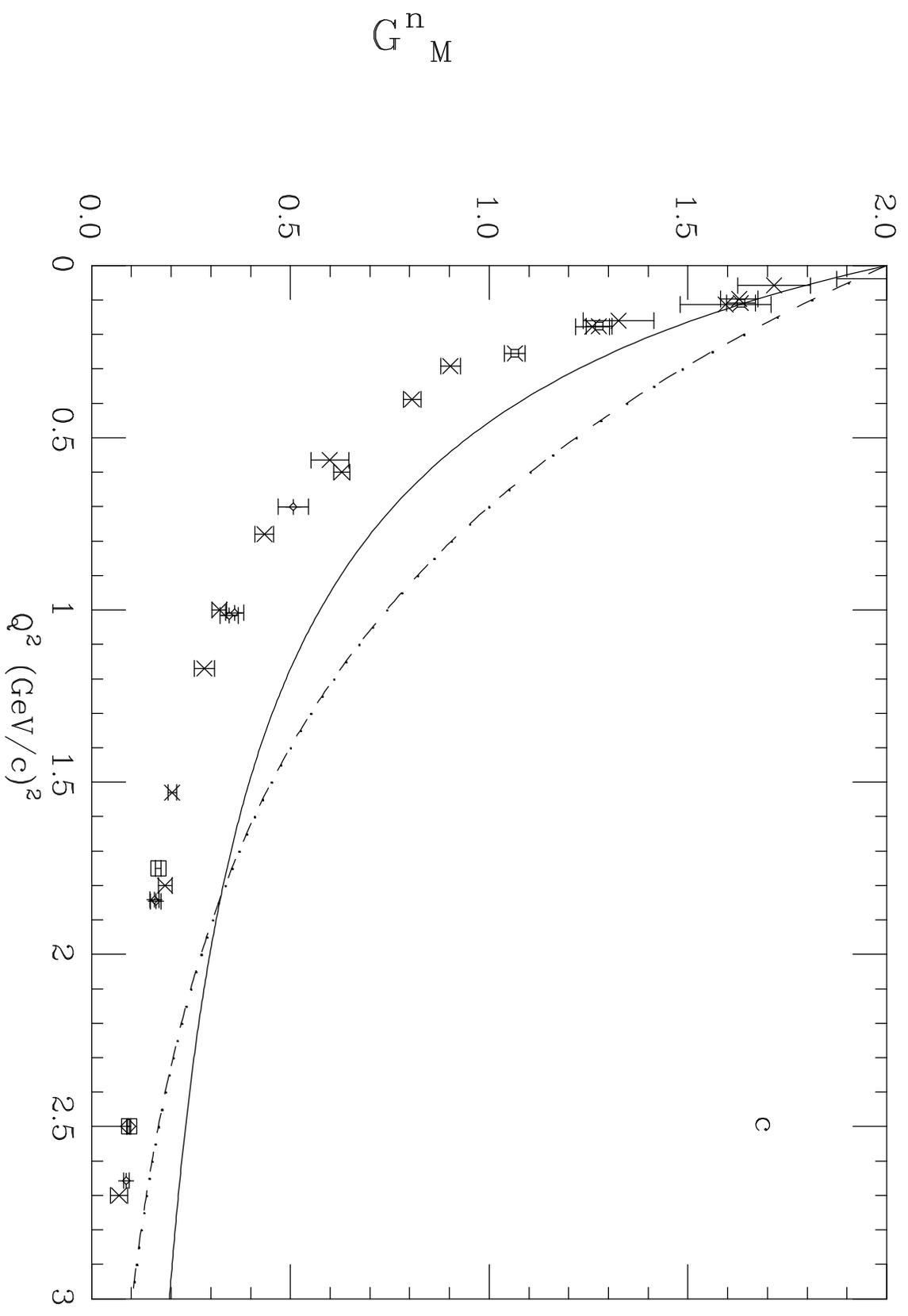
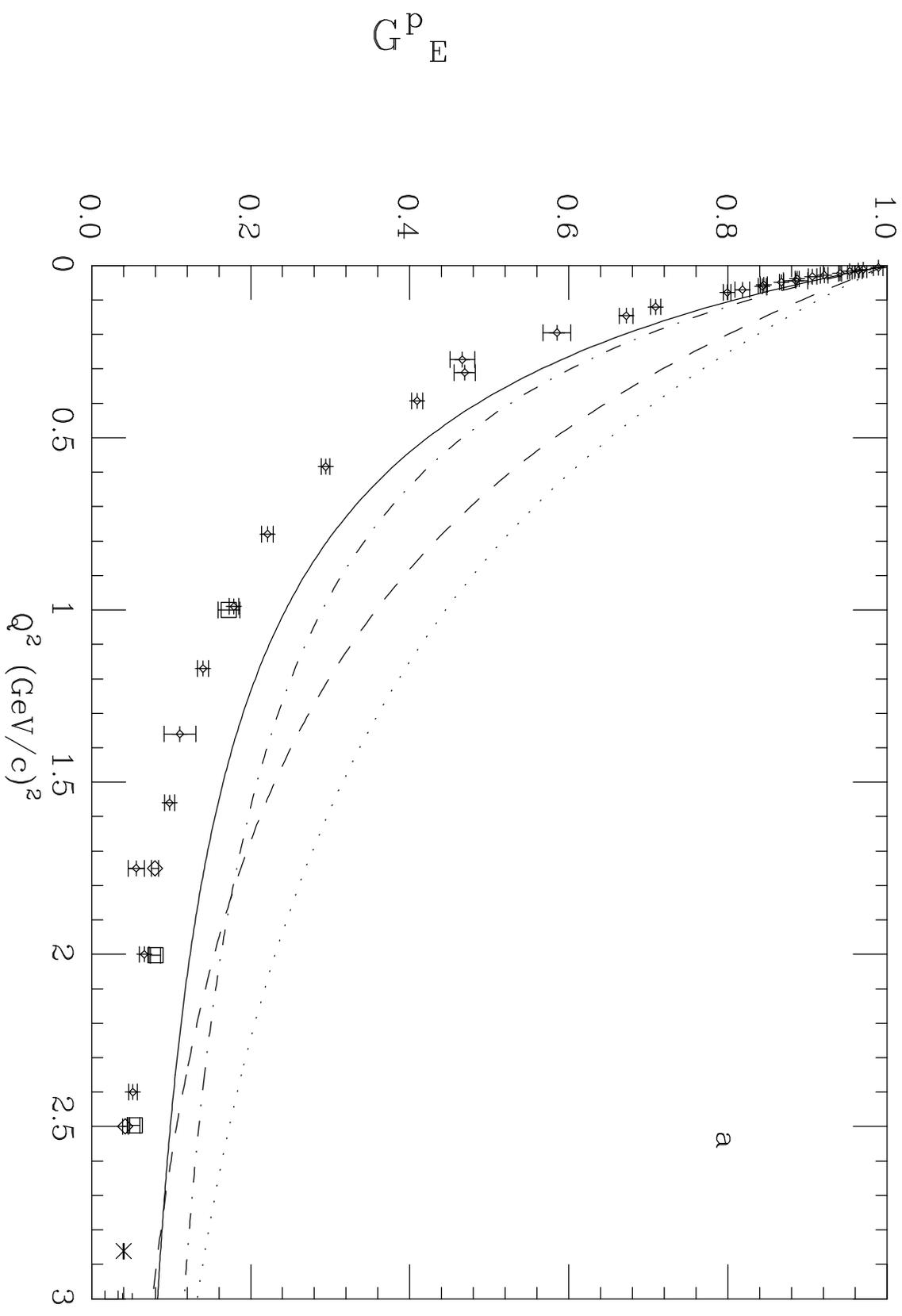
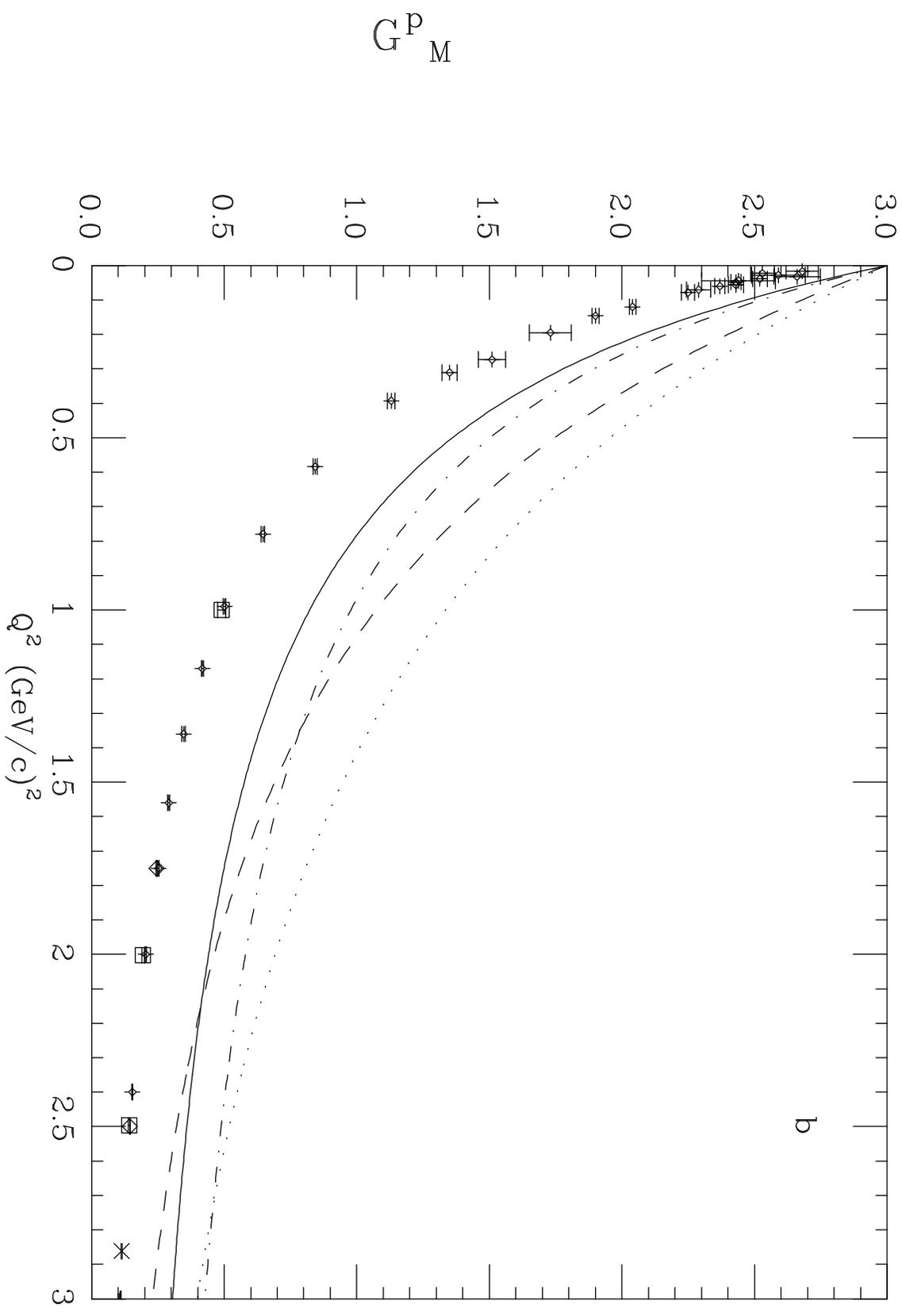
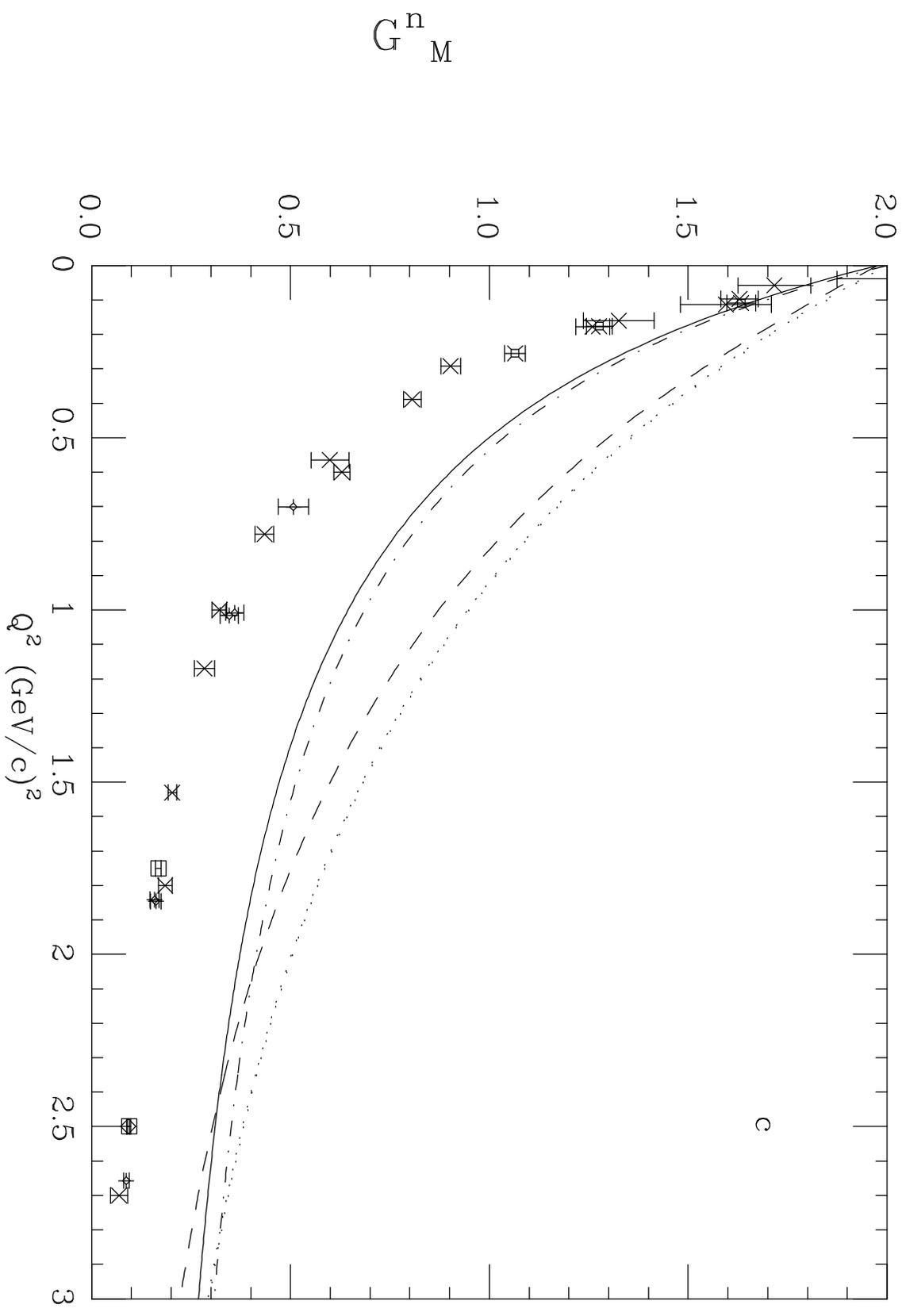
\end{document}